\documentclass[prl,twocolumn,superscriptaddress,showkeys,preprintnumbers,amsmath,amssymb,showemail]{revtex4-1}
\usepackage{epsfig}
\usepackage{graphics}
\usepackage{xfrac}
\usepackage[applemac]{inputenc}
\usepackage{color}
\usepackage[normalem]{ulem}

\begin{document}

\title{Local anodic oxidation on hydrogen-intercalated graphene layers: \\oxide composition analysis and role of the silicon carbide substrate}

\author{Francesco Colangelo}\affiliation{Laboratorio NEST - Scuola Normale Superiore $\&$ Istituto Nanoscienze - CNR, Piazza San Silvestro 12, I-56127 Pisa, Italy}\author{Vincenzo Piazza}\affiliation{Center for Nanotechnology Innovation@NEST - Istituto Italiano di Tecnologia, Piazza San Silvestro 12, I-56127 Pisa, Italy}\author{Camilla Coletti}\affiliation{Center for Nanotechnology Innovation@NEST - Istituto Italiano di Tecnologia, Piazza San Silvestro 12, I-56127 Pisa, Italy}\author{Stefano Roddaro}\affiliation{Laboratorio NEST - Scuola Normale Superiore $\&$ Istituto Nanoscienze - CNR, Piazza San Silvestro 12, I-56127 Pisa, Italy}\author{Fabio Beltram}\affiliation{Laboratorio NEST - Scuola Normale Superiore $\&$ Istituto Nanoscienze - CNR, Piazza San Silvestro 12, I-56127 Pisa, Italy}\author{Pasqualantonio Pingue}\email[Corresponding author: ]{pasqualantonio.pingue@sns.it}\affiliation{Laboratorio NEST - Scuola Normale Superiore $\&$ Istituto Nanoscienze - CNR, Piazza San Silvestro 12, I-56127 Pisa, Italy}
\begin{abstract}

We investigate nanoscale local anodic oxidation (LAO) on hydrogen-intercalated graphene grown by controlled sublimation of silicon carbide (SiC). Scanning probe microscopy (SPM) was used as a lithographic and characterization tool in order to investigate the local properties of the nanofabricated structures. The anomalous thickness observed after the graphene oxidation process is linked to the impact of LAO on the substrate. Micro-Raman ($\mu$-Raman) spectroscopy was employed to demonstrate the presence of two oxidation regimes depending on the applied bias. We show that partial and total etching of monolayer graphene can be achieved by tuning the bias voltage during LAO. Finally, a complete compositional characterization was achieved by scanning electron microscopy and energy dispersive spectroscopy (EDS).

\end{abstract}

\maketitle

\section{Introduction}

Starting from the pioneering work done by J. Dagata \emph{et al.} on silicon~\cite{Dagata}, spatially-resolved local anodic oxidation (LAO) induced by a scanning probe microscope (SPM) has been demonstrated on the nanometer scale on a variety of different substrates~\cite{Tseng}. The spectroscopic composition of the produced oxide patterns was also analyzed~\cite{LazzarinoPingueLAO, LazzarinoPadovaniLAO, Tello}. Recently, LAO was also successfully applied to graphene~\cite{NovoselovPNAS,NovoselovScience} for the fabrication of nanodevices, both using mechanically-exfoliated flakes as well as material obtained by SiC sublimation~\cite{PingueReview}. In these first reports, graphene oxidation was typically shown to lead to the local etching of the flake, following the production of carbon-based volatile compounds ~\cite{Rokhinson,NovoselovPSS}. Some researchers observed sometime the presence of "`insulating graphene"' in form of protruding regions during the LAO process, resulting in an increased roughness of the patterned regions ~\cite{Machida,NovoselovPSS}. These latter publications led to the conclusion that a graphene oxide (GO) layer was locally created by the LAO process. Conventionally, the oxide formation process is attributed to an electrochemical process directly involving C-O bond formation between graphene and decomposed OH$^-$ ions in the water meniscus that forms between tip and sample surfaces, although only few of these studies investigate specifically the structural and compositional properties of the oxidized graphene, e.g. employing Raman spectroscopy on regions patterned at various tip-sample voltages.

The impact of the bias voltage on the LAO process was addressed by various authors. The work of Masubuchi \textit{et al.} demonstrated that on a mechanically-exfoliated graphene lying on top of silicon dioxide (SiO$_2$) substrate the increase of tip-sample negative voltage can induce a correspondent increase in the oxidation level. The larger widths of the G and D peaks in the Raman spectrum \cite{FerrariRaman}, along with the diminished intensity of the 2D peak, with respect to bare graphene, were in this case indicators of the structural changes driven by SPM-induced oxidation (see SI on Masubuchi \textit{et al.} \cite{Machida2}). The composition of the oxidized regions formed on graphene deposited on SiO$_2$ was studied by $\mu$-Raman spectroscopy~\cite{Chuang}, showing in this case that the thickness of the produced nanostructures could not be solely attributed to GO formation: the coexistence of GO on the surface and sub-surface silicon oxide appears to be a better explanation for the authors. A complete study of the graphene composition after the SPM lithography was recently done by Byun \textit{et al.} ~\cite{Byun} and Chien \textit{et al.}~\cite{Hsiao-Mei Chien}, at negative and positive tip-sample voltages with different results. In the first case, by using $\mu$-Raman spectroscopy it was demonstrated that, at negative tip-to-sample voltages, a GO layer is produced, while at positive voltages a reversible hydrogenation of the graphene layer on SiO$_2$ occurs. In the second article, $\mu$-Raman and micro-X-ray photoelectron spectroscopy ($\mu$XPS) were employed on CVD-grown graphene deposited on SiO$_2$ and it was shown that protruding or recessed structures were formed indifferently by positive or by negative voltages and that bond reconstruction after oxidation is responsible for defect generation in both topographies.

The possibility to perform LAO was also demonstrated on a graphene layer obtained by SiC sublimation~\cite{Perez-Murano,Alaboson}. In one case, LAO was used to create insulating lines for both voltage polarities and oxidized regions were characterized by electric force microscopy (EFM)~\cite{Perez-Murano}. However, no compositional characterization of the fabricated structures was reported. A further study by Alaboson et al.~\cite{Alaboson} showed that, by etching the LAO structures obtained on graphene grown on SiC using hydrofluoric (HF) acid, a partial oxidation of the SiC substrate takes place that can  explain the anomalous thickness measured by SPM imaging. Also in this case, however, no compositional analysis of the oxidized regions was reported, and the authors only characterized the sample by exploiting selective wet etching and EFM on the SPM-fabricated structures. Therefore, while LAO directly performed on bare SiC substrate was already studied in the past~\cite{Lorenzoni,Xie} to the best of our knowledge,  no compositional data on graphene grown on SiC and oxidized by SPM via LAO technique are available at the moment in the literature.

In this article, we aim to fill this gap by presenting a study on the structural properties of nanostructures produced by LAO at negative tip-sample voltages, using mono- and bi-layer graphene obtained by sublimation of SiC and hydrogen intercalation process. This growth technique is particularly relevant for electronic applications thanks to the high-quality graphene that can be obtained on large scales and on semi-insulating substrates. In combination with LAO, this opens the way to the implementation of high-mobility nano-devices. In particular, in our work we focus on the oxide produced on a hydrogen-intercalated graphene sample, where the so-called ``buffer-layer" (the first carbon layer grown on a SiC substrate) is transformed into a standard mono-layer graphene by substitution and saturation with hydrogen of the Si-C covalent bonds created after the Si sublimation process~\cite{Coletti}.
 
\section{Experiment}

Graphene was grown by thermal decomposition of SiC(0001), as described in ref.~\cite{Starke}. Specifically, nominally on-axis SiC(0001) samples were first made atomically flat by hydrogen etching and then buffer-layer graphene was grown in Ar atmosphere at about 1350$^o$C. Hydrogen intercalation was performed as described in Ref.~\cite{Coletti} so that the grown buffer layer was turned into a "`quasi-free-standing"' monolayer graphene (QFMLG). Consistently with the literature~\cite{Goler}, SPM imaging correlated with $\mu$-Raman mapping show that the single layer graphene was found to be homogeneous in the inner area of the atomic terraces of SiC, while bilayer graphene was mainly detected at terrace edges. Moreover, the hydrogen intercalation process transforms the monolayer graphene into a ``quasi-free standing'' bi-layer graphene (QFBLG) as in the case of the QFMLG. In the following, we therefore report the study of LAO patterning on QFMLG and on QFBLG regions. 

\begin{figure}
\includegraphics[width=0.48\textwidth]{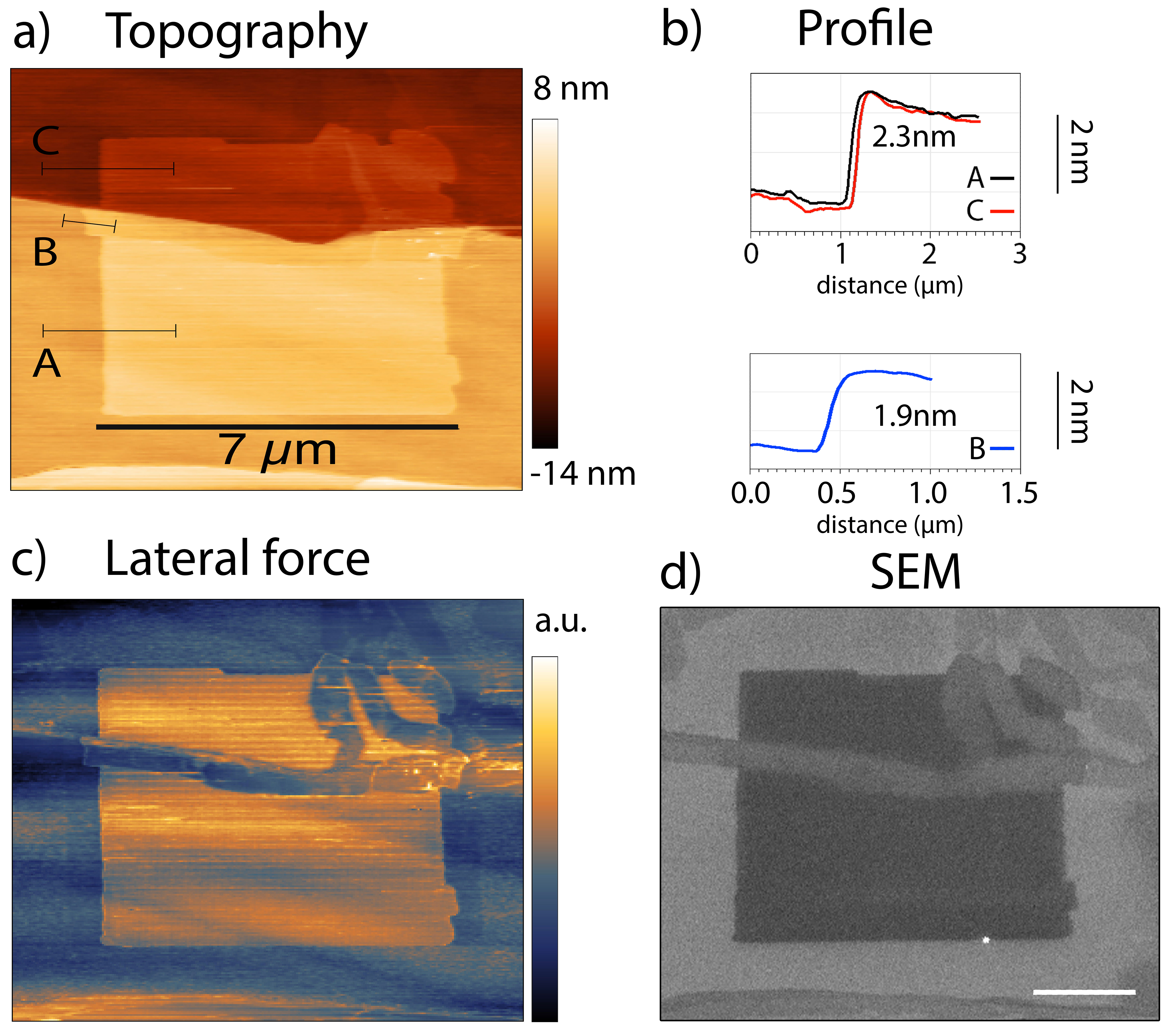}
\caption{Typical LAO structure produced on SiC substrate: (a) SPM topography;(b) SPM profile on the QFMLG (along lines A,C) and QFBLG (along segment B) oxidated regions; Lateral force image (c)  and SEM (d) of the same region are also shown.}
\label{fig1}
\end{figure}

LAO was performed in contact mode and in a humidity-controlled environment by employing a ``Caliber'' SPM system (Bruker, formerly Veeco) and standard p-doped Si tips coated in W$_2$C. During the process, the tip was grounded and the sample biased to a voltage $V_S$. All oxidations in this study were performed at negative tip-voltage bias, i.e. at positive $V_S$. The lithographic software allowed to define simple lines as well as more complex shapes with controllable tip-sample velocity, voltage and force in order to optimize the oxidation parameters. Topographic characterization was done in contact mode and in tapping mode therefore exploiting lateral-force microscopy (LFM) and phase-imaging techniques. Height measurements were always performed in contact mode to avoid artifacts~\cite{Biro}. The patterning parameters, such as ambient relative humidity (RH) and tip-sample velocity were optimized to maximize the reproducibility of the LAO process. They were then kept fixed during the experiments. We found a very good reproducibility with RH = 50\% and 0.5\,$\mu$m/s of tip-sample velocity. Various tip-to-sample bias were tested starting from +2 up to +10 Volts  (the maximum value available in our experimental setup), observing both a ``low-bias'' threshold ($\simeq$4-5 Volts) for the LAO process in our typical working conditions and an ``enhanced-oxidation'' process when the bias was set at +10 V and the right local boundary conditions (capillary water condensation, tip characteristics and scan speed) are fulfilled.

Figure~\ref{fig1}a reports a typical contact-mode image of a LAO structure fabricated on hydrogen-intercalated graphene on SiC, in correspondence of a terrace step. The squared bright area on top of the few-micron-wide terraces of the sample correspond to the region oxidized by a LAO process performed at $V_S=+10\,{\rm V}$. As visible in Fig.\ref{fig1}b and in the corresponding selected SPM profiles, oxidation induces a significant increase of the topographic thickness (on average, about $2-3\,{\rm nm}$). The structural modification of graphene is confirmed by lateral force microscopy (LFM) data. Friction is also typically found to strongly depend on the number of graphene layers~\cite{Lee}: for instance in Fig.\ref{fig1}c the darker region (smaller friction) appears to match the position of the oxidized QFBLG, while the brighter one (larger friction) to the oxidized QFMLG. In our investigation, this kind of SPM data was cross correlated with scanning electron microscopy (SEM), as visible in Fig.\ref{fig1}d, and with data obtained by $\mu$-Raman spectroscopy and by EDS, with the aim of better identifying the exact nature of the oxidized regions. Different $V_S$ values were used in order to probe the existence of different oxidation regimes. 
Raman characterization was performed using a standard Renishaw inVia system equipped with a $532\,{\rm nm}$ green laser and 100x objective lens.

\begin{figure}
\includegraphics[width=0.48\textwidth]{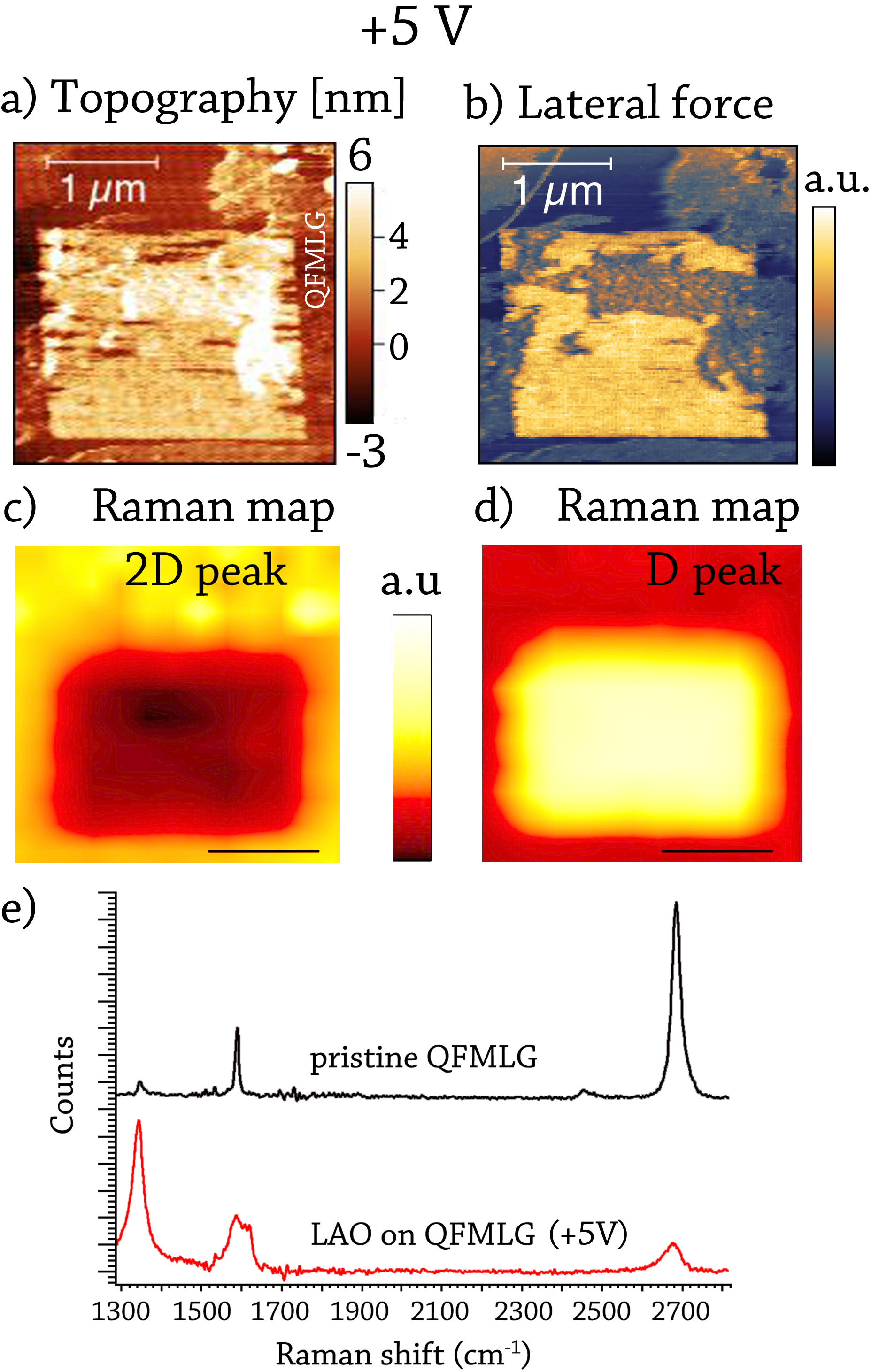}
\caption{SPM image of the rectangular shaped structures obtained by LAO process at $V_S=+5\,{\rm V}$. (a) SPM topographic image. (b) Lateral force image of the same region; $\mu$-Ramam map of the 2D (c) and D (d) peaks. (e) Comparison between a complete Raman spectrum acquired on pristine QFMLG (black) and oxidized QFMLG region (red).}
\label{fig2}
\end{figure}

\begin{figure}
\includegraphics[width=0.48\textwidth]{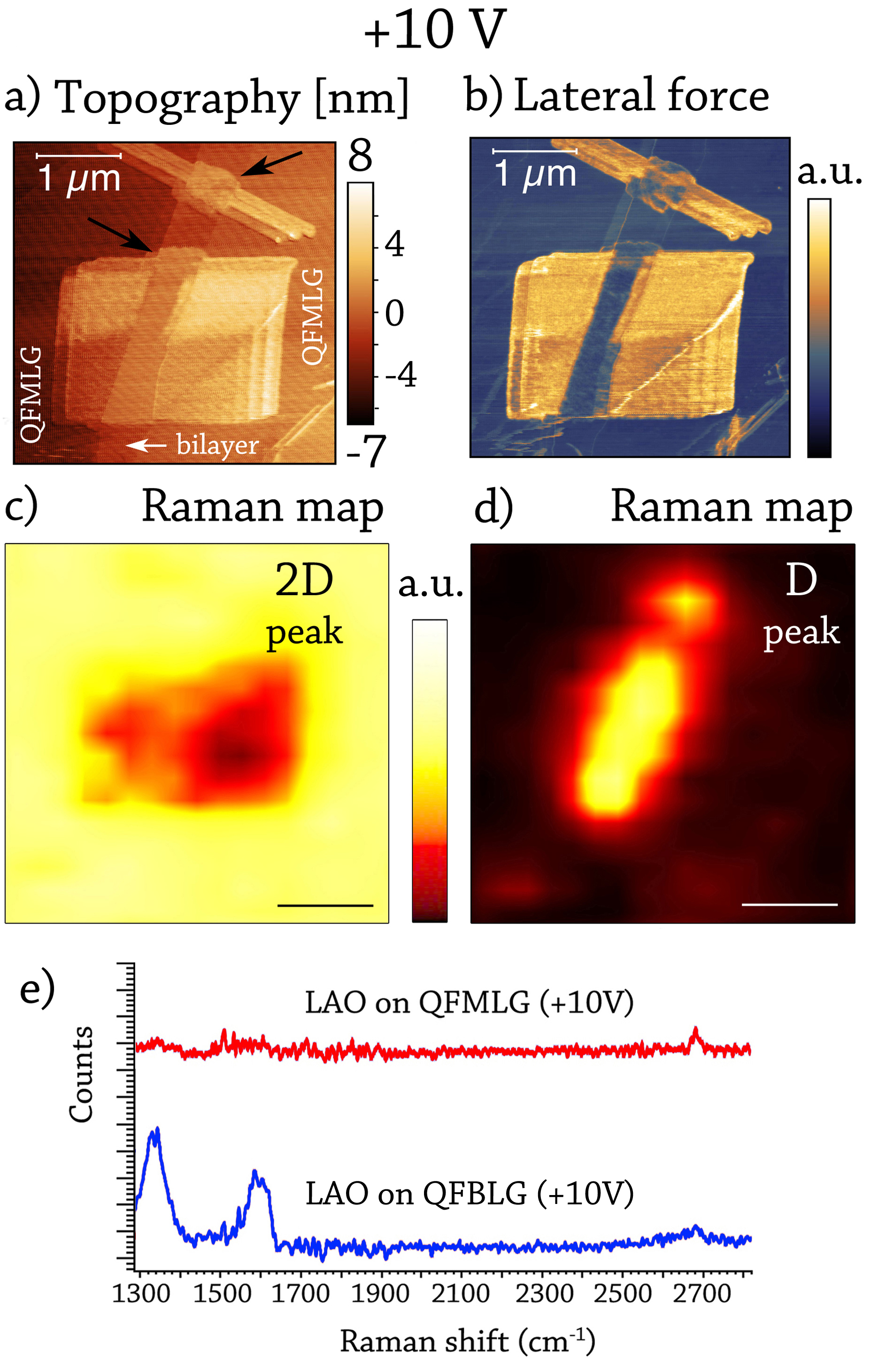}
\caption{SPM image of a rectangular shaped structures obtained by LAO process at $V_S=+10\,{\rm V}$. (a) SPM topographic image. (b) Lateral force image of the same region. A ``double-tip'' artifact is present in these images}. $\mu$-Ramam map of the 2D (c) and D (d) peaks. (e) Comparison between a complete Raman spectrum acquired on oxidized QFMLG region (red) and oxidized QFBLG (blue).
\label{fig3}
\end{figure}

\section{Experimental Results and Discussion}
 
Figures ~\ref{fig2} and ~\ref{fig3} show SPM images of two square-shaped structures produced by LAO on graphene in two different oxidation regimes. The square in Fig.~\ref{fig2}a was obtained by applying a bias voltage $V_S=+5\,{\rm V}$ while the square reported in Fig.~\ref{fig3}a was produced by applying $V_S=+10\,{\rm V}$ bias. These images show the two observed regimes (that we named ``standard-'' and ``enhanced-oxidation'') of the LAO process. The two were separately studied in order to analyze their characteristic properties. The protruding oxide structures have an average height of the order of $2-3\,{\rm nm}$. As already evidenced, this value is very large compared to graphene thickness and straightforwardly suggests that the substrate play a role in the process. Figures~\ref{fig2}b and Fig.~\ref{fig3}b show lateral-force images of the oxidized regions. It can be noticed that these show a different contrast with respect to the non-oxidized ones, indicating a modification of the surface roughness. More in detail, the friction value of the oxide produced on the QFMLG layer is higher with respect to that typically observed on pristine graphene, as can be expected from the fact that the roughness of its surface increases during the LAO process. This phenomenon was already observed in previous experiments performed on graphene flakes deposited on SiO$_2$ substrate~\cite{Byun}. On the other hand, LFM imaging on the oxidized region of QFBLG (see Fig.~\ref{fig1}c and ~\ref{fig3}b) shows a contrast similar to that of pristine graphene. Moreover, the QFBLG seems to have a different oxidation reactivity with respect to the QFMLG, as demonstrated by the areas protruding from the square edges, highlighted by the black arrows in Fig.~\ref{fig3}. This finding was already reported by another group in the case of LAO on non-hydrogenated graphene grown on SiC~\cite{Alaboson}. Therefore, the role of hydrogen intercalation does not seem to be relevant from this point of view. 

Starting from the observation of this anomalous thickness of the oxidized graphene regions, and in order to gain more information about the chemical nature of these SPM-fabricated regions, we employed $\mu-$Raman imaging and EDS.
We mapped the 2D and the D Raman-band intensities integrating the spectra from 2600 to 2800 cm$^{-1}$ and from 1300 to 1400 cm$^{-1}$ respectively.
Raman maps for oxidization using a $V_S=+5\,{\rm V}$ bias are reported in Fig.~\ref{fig2}c) and d).  The 2D-band signal was found to be weaker on the LAO-modified zone than the surrounding pristine QFMLG region(Fig.~\ref{fig2}c). On the contrary, in the oxidized square a significant signal appears in correspondence of the D Raman peak (Fig.~\ref{fig2}d) demonstrating the presence of defective graphene. As a reference, we report in Fig.~\ref{fig2}e the Raman spectra collected at the center of the square and on pristine QFMLG after subtraction of the SiC Raman background signal. The intensity ratio of the D and G Raman peaks is $I(\mathrm{D})/I(\mathrm{G})\simeq2.8$, the FWHM of G peak is about 50\,$\mathrm{cm}^{-1}$, $I(\mathrm{2D})/I(\mathrm{D})\simeq0.2$ and $I(\mathrm{2D})/I(\mathrm{G})\simeq0.6$. This indicate that the LAO modified graphene has a larger amount of defects (on average, one every $3-5\,{\rm nm}$), as indicated by ref.~\cite{Raman01}.
Moreover, as discussed in ref.~\cite{Raman02}, from the intensity ratio of D and D' Raman peaks it is possible to assess the nature of the defects: in our case the ratio $I(\mathrm{D})/I(\mathrm{D'})\simeq6$ suggests that the defects are ``\textit{vacancy}-like''. This conclusion is also supported by a direct comparison of our data with the Raman spectra reported by Childres \textit{et al.} in ref.~\cite{Raman03}. In the cited work authors studied how the Raman spectrum evolves after consecutive steps of oxygen plasma etching. After about 15-20 etch steps of 1 s each, the Raman spectrum results very similar in terms of peak intensity ratio ($I(\mathrm{D})/I(\mathrm{G})$ and $I(\mathrm{2D})/I(\mathrm{G})$) to the one collected in our LAO processed region.

In light of these facts, we conclude that the impact of the LAO process on QFMLG in this ``standard-oxidation'' regime is very similar to a partial {\em etching} of the graphene surface. This is consistent with a picture where the strong electric field present between the tip and the sample surface produces free radicals that damage the graphene monolayer and at the same time oxidize the SiC below the QFMLG. Thus, the final result is a thick topographic protrusion, as discussed above and confirmed by other groups~\cite{NovoselovPSS, Machida} in experiments done on graphene deposited on $SiO_2$. This ``partial etching'' picture of the LAO process is also confirmed by other groups with data obtained on single- and multi-layer graphene region deposited on SiO$_2$ substrate and oxidated in O$_2$/Ar atmosphere at various temperatures. In these cases, AFM imaging shows the presence of etch pits (having a diameter of $\simeq20 nm$) on the graphene surface~\cite{Flynn, Liu}. 

Interestingly, when the LAO process is performed at an increased bias voltage ($V_S=+10\,{\rm V}$) similar modifications (as shown in Fig.~\ref{fig3}a) were observed, but an even more drastic change of the lateral extension of the oxide layer and of the Raman properties on the QFMLG layer was obtained if the ``enhanced-oxidation'' regime is involved. Figure~\ref{fig3}c and Fig.~\ref{fig3}d report the maps of the 2D and the D Raman bands of the oxidized square region. In both maps the Raman signal is strongly depressed on the QFMLG oxidized area. Indeed, as visible in Fig.~\ref{fig3}e, no graphene-related peak is visible in the full Raman spectrum of the oxidized region. We thus conclude that, for LAO in the ``enhanced-oxidation regime'', the QFMLG is completely removed probably due to the formation of volatile carbon compounds (e.g. CO or CO$_2$ molecules) during the lithographic process.

\begin{figure}
\includegraphics[width=0.48\textwidth]{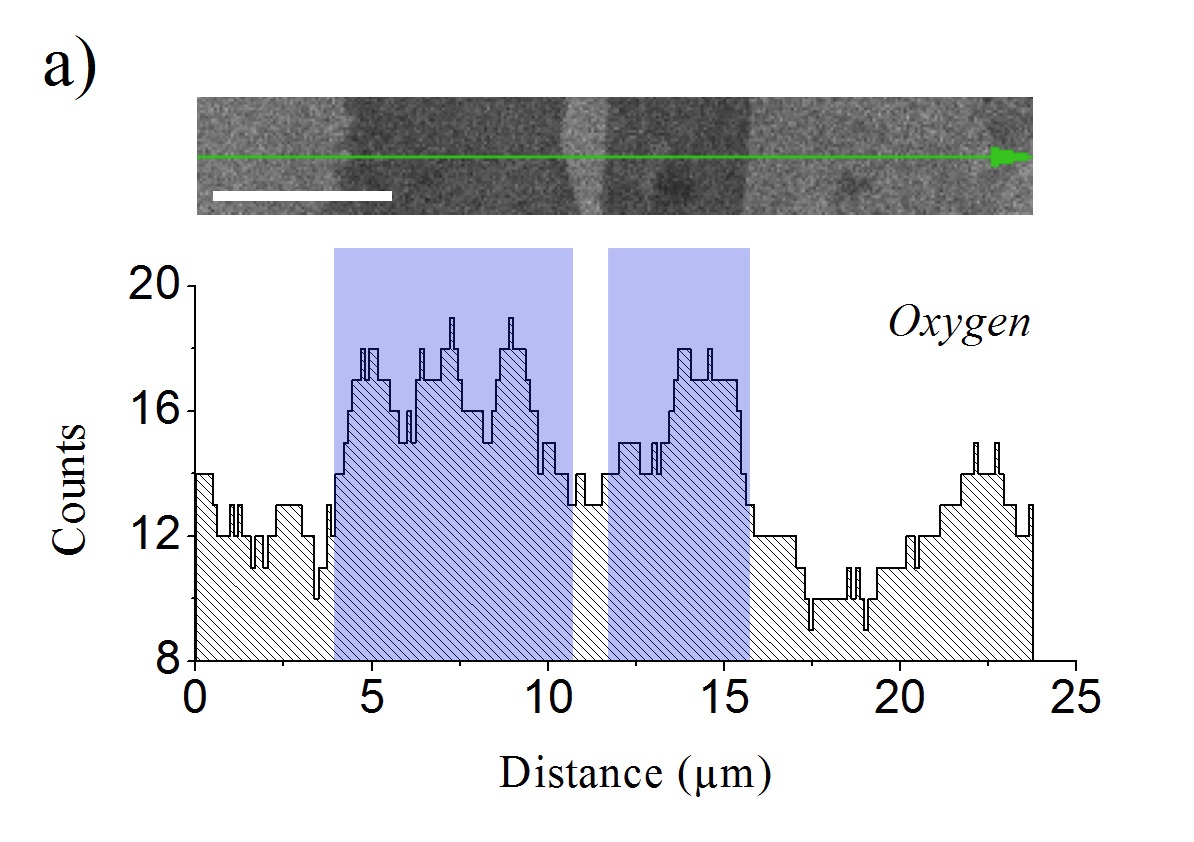}
\includegraphics[width=0.48\textwidth]{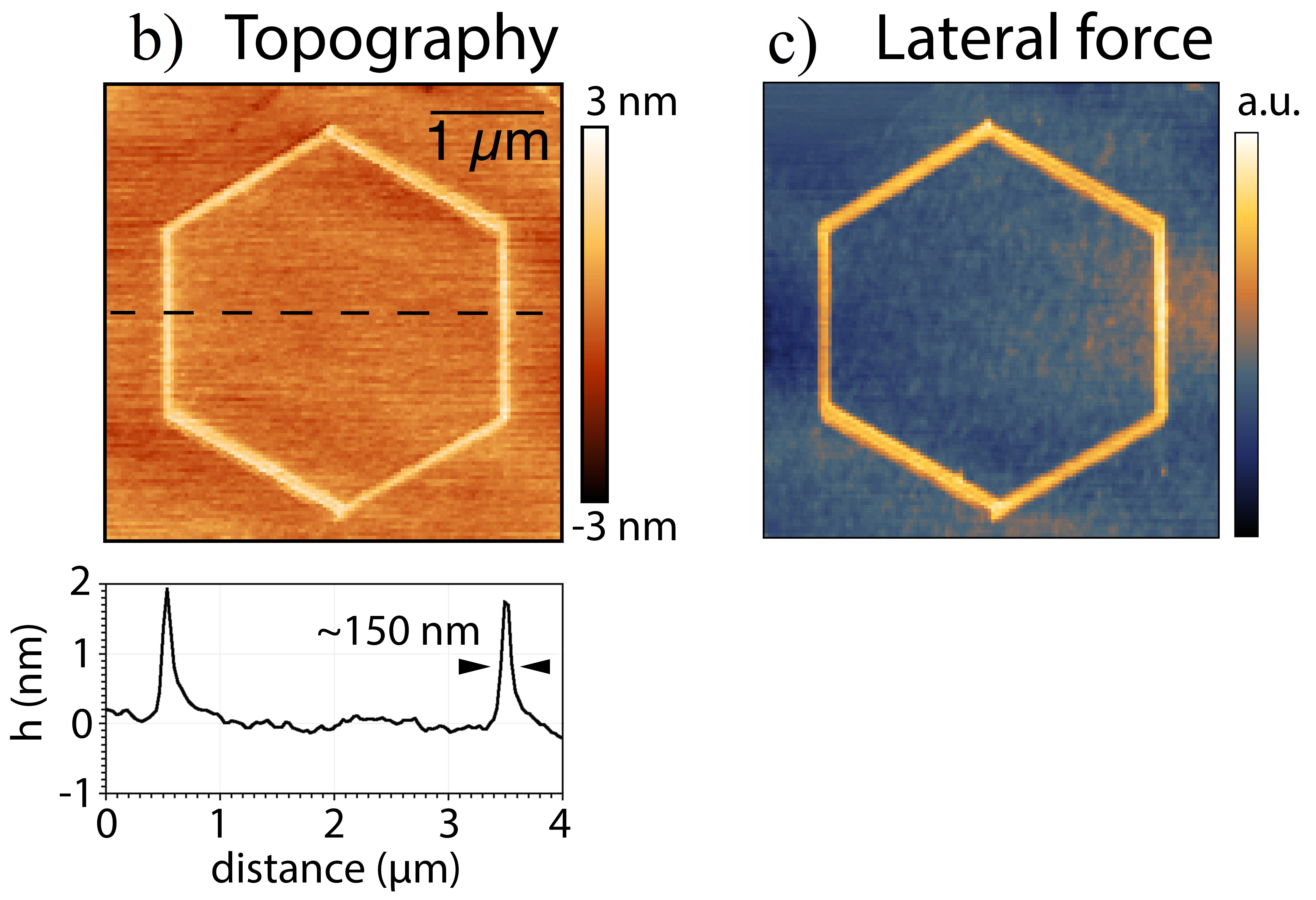}
\caption{(a) Energy-dispersive X-ray spectroscopy (EDS) of graphene regions subject to LAO in the ``enhanced-oxidation'' regime. The oxygen content of the sample is analized along the cross-section indicated by the green arrow in overlay in the SEM picture at the top of the panel, where darker regions result form the LAO process. An excess of oxygen is detected in correspondence to the stripes (shaded blue regions of the graph), confirming the local oxidation of the sample. (b) AFM topography and local profile of a patterned hexagon obtained by LAO on QFMLG and (c) corresponding LFM image.}
\label{fig4}
\end{figure}

Further interesting evidence emerge from the Raman maps when comparing regions corresponding to oxidized QFBLG in comparison with oxidized QFMLG (see Fig.~\ref{fig3}c,d). Indeed, in QFBLG the typical Raman bands of graphene are also depressed, but not completely absent. A spectrum collected in the QFBLG region is reported in Fig.~\ref{fig3}e (red curve): a broad G band is still present, together with a pronounced D peak and a very broad and depressed 2D peak. The D peak map in Fig.~\ref{fig3}d indicates that the oxidized QFBLG is spatially homogeneous, consistently with LFM data (Fig.~\ref{fig3}b). We thus conclude that this region most likely consists of highly-disordered graphene oxide and is not completely etched as in the case of the oxidized QFMLG.  
These data were also compared with EDS measurements in order to confirm the nature of the regions subjected to LAO at $V_S=+10\,{\rm V}$ and extract information about the local composition of the sample. From this analysis and all the acquired spectra, we did not observed the presence of any contaminants in the patterned region (e.g. metallic salts) but only the presence of Si, C,and O elements. Fig.~\ref{fig4}a shows the presence of excess oxygen in the region subject to the LAO process as expected. Present accuracy in the determination of the local oxygen content does not allow us to clearly identify the likely oxidation of the SiC substrate, but does confirm that our procedure leads to a local oxidation of the graphene-SiC sample.

LAO technique allows to easily produce more complex patterns on top of the graphene/SiC layer, exploiting the AFM lithographic and positioning capabilities (see as example the hexagonal shape patterned at +8 V on QFMLG in Fig.~\ref{fig4}b,c) and opening in this way the route toward a SPM-based nanofabrication of functional devices on this material. The possibility to tune the selective oxidation of different graphene layers represents one of the interesting developments of this technique on semi-insulating substrates.

\section{Conclusion}

We performed LAO on graphene layers grown by sublimation of SiC and intercalated by hydrogen to form QFMLG and QFBLG, observing by SPM imaging that a thick oxide layer is formed on top of the surface for different negative tip-to-sample bias. We studied the composition of the produced oxide by $\mu$-Raman spectroscopy and imaging showing the presence of two different regimes in LAO oxidation process: (i) a ``standard-oxidation'' regime that triggers the local and partial etching of the graphene;  (ii) an ``enhanced-oxidation'' regime that completely destroys the QFMLG graphene structure and yields high-disorder graphene on the QFBLG region which reacts differently.
$\mu$-Raman mapping, the presence of excess oxygen content in the patterned region (as evidenced by EDS imaging), and the observed high-protruding topography (as also found in literature on LAO process applied to bare SiC~\cite{Lorenzoni,Xie}) implies that LAO process drives the oxidation of the underlying SiC substrate together with the graphene top layer. This leads to the  formation of the observed `bumps' on the surface. A comparison with similar experiments performed on SiC substrate with no intercalation during the graphene growing step ~\cite{Alaboson} suggests that the presence of hydrogen does not play a crucial role in the observed oxidation results. 

These findings lay the ground for understanding LAO of graphene and for the optimization of SPM-based patterning of high-quality graphene layers grown on SiC. We believe these results will play an important role for the definition of viable protocols for the fabrication of graphene-based nanoelectronic devices on semi-insulating substrates.


\begin{references}

\bibitem{Dagata} J.A. Dagata, J. Schneir, H.H. Harary, C.J. Evans, M.T. Postek, and J. Benett, Appl. Phys. Lett. 56, 2001 (1990);

\bibitem{Tseng} For recent reviews, see: ``Tip-Based Nanofabrication: Fundamentals and Applications'', Ampere Tseng Ed., Springer (2011); ``Advanced scanning probe lithography'', by R. Garcia, A.W. Knoll and E. Riedo, Nature Nanotechnology, 9, 577 (2014); ``Nanoscale Materials Patterning by Local Electrochemical Lithography'' by He Liu, S. Hoeppener and U.S. Schubert, Adv. Eng. Mat. 18 (6), 890 (2016).

\bibitem{LazzarinoPingueLAO} M. Lazzarino, S. Heun, B. Ressel, K. C. Prince, P. Pingue, and C. Ascoli, Appl. Phys. Lett. 81, 2842 (2002);

\bibitem{LazzarinoPadovaniLAO} M. Lazzarino, M. Padovani, G. Mori, L. Sorba, M. Fanetti, M. Sancrotti, Chemical Physics Letters 402, 155-159 (2005);

\bibitem{Tello} M. Tello, R. Garcia, J. A. Mart\'in-Gago, N. F. Martinez, M. S. Martin-Gonzalez, L. Aballe, A. Baranov and L. Gregoratti, Adv. Mater. 17, 1480 (2005);

\bibitem{NovoselovPNAS} K. S. Novoselov, D. Jiang, F. Schedin, T. J. Booth, V. V. Khotkevich, S. V. Morozov, and A. K. Geim, PNAS  vol. 102 (30), 10451-10453, (2005);

\bibitem{NovoselovScience} K. S. Novoselov, A. K. Geim, S. V. Morozov, D. Jiang, M.I. Katsnelson, I. V. Grigorieva, S.V. Dubonos, and A.A. Firsov, Nature 438, 197-200 (2005);

\bibitem{PingueReview} Review P. Pingue, Chapter on `Tip-based Nanofabrication: Fundamentals and Apllications', Ampere Tseng Ed., Springer (2011);

\bibitem{Rokhinson} L. Weng, L. Zhang, Y.P. Chen and L.P. Rokhinson, Appl. Phys. Lett. 93, 093107 (2008);

\bibitem{NovoselovPSS} S. Neubeck, F. Freitag, R. Yang and K. S. Novoselov, Physica Status Solidi (b), 247(11-12), 2904-2908 (2010);

\bibitem{Machida2} S. Masubuchi, M. Arai and T. Machida, Nano letters, 11(11), 4542-4546 (2011);

\bibitem{Machida} S. Masubuchi, M. Ono, K. Yoshida, K. Hirakawa, and T. Machida, Appl. Phys. Lett. 94, 082107 (2009);

\bibitem{FerrariRaman} A. C. Ferrari, J. C. Meyer, V. Scardaci, C. Casiraghi, M. Lazzeri, F. Mauri, S. Piscanec, D. Jiang, K. S. Novoselov, S. Roth, and A. K. Geim, Phys. Rev. Lett. 97, 187401 (2006)

\bibitem{Chuang} Min-Chiang Chuang, Hsiao-Mei Chien, Yuan-Hong Chain, Gou-Chung Chi, Sheng-Wei Lee, and Wei Yen Woon, Carbon 54, 336-342 (2013);

\bibitem{Byun} Ik-Su Byun, Duhee Yoon, Jin Sik Choi, Inrok Hwang, Duk Hyun Lee, Mi Jung Lee, Tomoji Kawai, Young-Woo Son, Quanxi Jia, Hyeonsik Cheong, and Bae Ho Park, ACS Nano vol.5(8), 6417-6424 (2011);

\bibitem{Hsiao-Mei Chien} Hsiao-Mei Chien, Min-Chiang Chuang, Hung-Chieh Tsai, Hung-Wei Shiu, Lo-Yueh Chang, Chia-Hao Chen, Sheng-Wei Lee, Jonathon David White, Wei-Yen Woon, Carbon 80, 318 (2014);

\bibitem{Perez-Murano} G. Rius, N. Camara, P. Godignon, and F. P\'erez-Murano, and N. Mestres, J. Vac. Sci. Technol. B 27, 3149-3152 (2009); 

\bibitem{Alaboson} Justice M. P. Alaboson , Qing Hua Wang , Joshua A. Kellar , Joohee Park , Jeffrey W. Elam ,  Michael J. Pellin, and  Mark C. Hersam, Adv. Mater. 23, 2181-2184 (2011);

\bibitem{Lorenzoni} M. Lorenzoni and B. Torre, Appl. Phys. Lett. 103, 163109 (2013);

\bibitem{Xie} X.N. Xie, H.J. Chung, C.H. Sow, and A.T.S. Wee, Appl. Phys. Lett., vol. 84 (24), 4914-4916 (2004);

\bibitem{Coletti} C. Riedl, C. Coletti, T. Iwasaki, A. A. Zakharov, and U. Starke, Phys. Rev. Lett. 103, 246804 (2009);

\bibitem{Starke} U Starke, S Forti, KV Emtsev, C Coletti, Mrs Bulletin 37 (12), 1177-1186 (2012);

\bibitem{Goler} S. Goler, C. Coletti, V. Piazza, P. Pingue, F. Colangelo, V. Pellegrini, K.V. Emtsev, S. Forti, U. Starke, F. Beltram, S. Heun, Carbon 51, 249-254 (2013); 

\bibitem{Biro} Nemes-Incze, P., Osv\'ath, Z., Kamar\'as, K. and Bir\'o, L. P., Carbon 46, 1435 (2008);

\bibitem{Lee} C. Lee, Q. Li, W. Kalb, Xin-Zhou Liu, H. Berger, R.W. Carpick, J. Hone, Science 328, 76 (2010);

\bibitem{Raman01} L. G. Can\c{c}ado, A. Jorio, E. H. Martins Ferreira, F. Stavale, C. A. Achete, R. B. Capaz, M. V. O. Moutinho, A. Lombardo, T. S. Kulmala and A. C. Ferrari, Nano Letters, vol. 11 (8), 3190-3196 (2011);

\bibitem{Raman02} A. Eckmann, A. Felten, A. Mishchenko, L. Britnell, R. Krupke, K. S. Novoselov and Cinzia Casiraghi, Nano Letters, vol. 12 (8), 3925-3930 (2012);

\bibitem{Raman03} I. Childres, L. A. Jauregui, J. Tian and Y. P. Chen, New Journal of Physics, 13(2), 025008 (2011);

\bibitem{Flynn} L. Liu, S. Ryu, M.R. Tomasik, E. Stolyarova, N. Jung, M.S. Hybertsen, M.L. Steigerwald, L.E. Brus, and G.W. Flynn, Nano Letters 7 (8) 1965 (2008).

\bibitem{Liu} S.P. Surwade, Z. Li, and H. Liu, The J. of Phys. Chem. C 116 (38), 20600 (2012).

\end{references}
\end{document}